\documentclass[floats, eqnum, showpacs, nofootinbib, twocolumn, eqsecnum]{revtex4-1}

\usepackage{color,graphicx}
\usepackage{amsfonts}
\usepackage{amssymb}
\usepackage{amsmath}

\begin{document}

\title{Extended source effect on microlensing light curves by an Ellis wormhole}
\author{Naoki Tsukamoto}
\email{tsukamoto@rikkyo.ac.jp}
\author{Yungui Gong}
\email{yggong@hust.edu.cn}
\affiliation{School of Physics, Huazhong University of Science and Technology, Wuhan, Hubei 430074, China}

\begin{abstract}
We can survey an Ellis wormhole which is the simplest Morris-Thorne wormhole in our galaxy with microlensing. 
The light curve of a point source microlensed by the Ellis wormhole shows approximately $4\%$ demagnification 
while the total magnification of images lensed by a Schwarzschild lens is always larger than unity.
We investigate an extended source effect on the light curves microlensed by the Ellis wormhole.
We show that the depth of the gutter of the light curves of an extended source is smaller than the one of a point source 
since the magnified part of the extended source cancels the demagnified part out.
We can, however, distinguish between the light curves of the extended source microlensed by the Ellis wormhole 
and the ones by the Schwarzschild lens in their shapes
even if the size of the source is a few times larger than the size of an Einstein ring on a source plane. 
If the relative velocity of a star with the radius of $10^6$~km at $8$~kpc in the bulge of our galaxy against an observer-lens system is smaller than $10$~km/s on a source plane,
we can detect microlensing of the star lensed by the Ellis wormhole with the throat radius of $1$~km at $4$~kpc. 
\end{abstract}

\maketitle

\section{Introduction}
General relativity permits spacetimes with nontrivial topological structures such as a wormhole spacetime 
which is the nonvacuume solution of Einstein equations~\cite{Visser_1995}.
Gravitational lensing~\cite{Schneider_Ehlers_Falco_1992,Petters_Levine_Wambsganss_2001,Schneider_Kochanek_Wambsganss_2006,Perlick_2004_Living_Rev} 
is a good tool to survey for not only dark and massive objects but also wormholes.
After pioneer works in gravitational lensing by a wormhole by Kim and Cho~\cite{Kim_Cho_1994} and Cramer~\textit{et al.}~\cite{Cramer:1994qj},
gravitational lensing by wormholes with a positive Arnowitt-Deser-Misner (ADM) mass~~\cite{Nandi_Zhang_Zakharov_2006,
Rahaman:2007am,Tejeiro_Larranaga_2012,Kuhfittig:2015sta,Sajadi:2016hko,Nandi:2016ccg,Nandi:2016uzg,Tsukamoto:2016zdu,Jusufi:2017mav,Shaikh:2017zfl,Goulart:2017iko}, 
with a vanishing ADM mass~\cite{Chetouani_Clement_1984,Perlick_2004_Phys_Rev_D,Nandi_Zhang_Zakharov_2006,Muller:2008zza,
Abe_2010,Toki_Kitamura_Asada_Abe_2011,Tsukamoto_Harada_Yajima_2012,Yoo_Harada_Tsukamoto_2013,
Takahashi_Asada_2013,Tsukamoto_Harada_2013,Izumi_2013,Nakajima:2014nba,Bozza:2015haa,Bozza:2015wbw,Tsukamoto:2016qro,Tsukamoto:2016jzh,Tsukamoto:2016zdu,Lukmanova_2016,Tsukamoto:2017edq,Jusufi:2017gyu,Jusufi:2017vta,Bozza:2017dkv,Asada:2017vxl}, 
and with a negative ADM mass~\cite{Cramer:1994qj,Torres:1998xd,Takahashi_Asada_2013,Shaikh:2017zfl} were investigated.

An Ellis wormhole~\cite{Ellis_1973,Bronnikov_1973} which is the solution of Einstein equation with a phantom scalar field 
is the simplest and earliest Morris-Thorne wormhole~\cite{Morris_Thorne_1988}. 
Instability of the Ellis wormhole was reported in Refs.~\cite{Shinkai_Hayward_2002}, contrary to a conclusion of an earlier work~\cite{Armendariz-Picon_2002}.
The uniqueness theorem of the Ellis wormhole in the Einstein-phantom scalar field theory has been given~\cite{Yazadjiev:2017twg}. 

Wormhole solutions without the phantom scalar field and with the other exotic matters as the sources which have the same metric of the Ellis wormhole 
were also obtained in Refs.~\cite{Kar:2002xa,Das:2005un,Shatskiy:2008us,Novikov:2012uj,Myrzakulov:2015kda}.
Stability of the wormholes depends on both the metric and the sources.
The fact that a wormhole by an electrically charged dust with negative energy density 
as the source which is the same as the metric of the Ellis wormhole~\cite{Shatskiy:2008us,Novikov:2012uj}
is linearly stable against both spherical and axial perturbations was found by Bronnikov~\textit{et al.} in 2013~\cite{Bronnikov:2013coa}.
The wormhole solution might be the first example of stable wormholes without a thin shell in general relativity.
The quasinormal mode of the wormhole was also investigated in Ref.~\cite{Konoplya:2016hmd}. 

The deflection angle of a light ray in the Ellis wormhole spacetime was obtained by~Chetouani and Cl\'{e}ment for the first time in 1984~\cite{Chetouani_Clement_1984}
and it was reexamined several times~\cite{Nandi_Zhang_Zakharov_2006,Muller:2008zza,Tsukamoto:2016qro,Tsukamoto:2016jzh,Jusufi:2017gyu}.
The upper bound of the number density of the Ellis wormhole with the radius of a throat $10\leq a \leq 10^4$~pc was given 
as $10^{-4}h^3 \mathrm{Mpc}^{-3}$~\cite{Takahashi_Asada_2013} 
by the strong lensing of quasars in the data of the Sloan Digital Sky Survey Quasar Lens Search~\cite{Inada_Oguri_Shin_et_al_2012}
and the one with $a\sim 1$~cm was given as $10^{-9} \mathrm{AU}^{-3}$~\cite{Yoo_Harada_Tsukamoto_2013}
by the femtolensing of gamma-ray bursts~\cite{Barnacka_Glicenstein_Moderski_2012} 
in the data of the Fermi Gamma-Ray Burst Monitor~\cite{Meegan_Lichti_Bhat_et_al_2009}.
In the Ellis wormhole spacetime, the shear~\cite{Izumi_2013} and the time delay of lensed images~\cite{Nakajima:2014nba}, 
retrolensing~\cite{Tsukamoto:2016zdu,Tsukamoto:2017edq}, 
gravitational lensing of light rays passing through a throat~\cite{Perlick_2004_Phys_Rev_D,Tsukamoto:2016zdu}, 
Einstein rings~\cite{Tsukamoto_Harada_Yajima_2012}, 
microlensing~\cite{Abe_2010,Tsukamoto:2016zdu,Lukmanova_2016,Tsukamoto:2016qro},
astrometric image centroid displacements~\cite{Toki_Kitamura_Asada_Abe_2011}, 
binary lenses~\cite{Bozza:2015wbw}, 
and so on~\cite{Muller_2004,Tsukamoto_Harada_2013,Yoo_Harada_Tsukamoto_2013,Tsukamoto:2014swa,Ohgami:2015nra,Ohgami:2016iqm,Perlick:2015vta,Bozza:2015haa} were investigated.
(See Refs.~\cite{Tsukamoto_Harada_2013,Kitamura_Nakajima_Asada_2013,Izumi_2013,Nakajima:2014nba,Bozza:2015haa,Bozza:2015wbw,Bozza:2017dkv,Asada:2017vxl} 
for exotic lens objects with a gravitational potential which is asymptotically proportional to $1/r^n$, where $n$ is a positive number. 
An Ellis wormhole and a Schwarzschild lens have a gravitational potential with $n=2$ and $n=1$, respectively.) 

It is well known that the total magnification of images lensed by a positive mass is always larger than unity~\cite{Petters_Levine_Wambsganss_2001}.
We cannot, however, apply the magnification theorem for a Ellis wormhole lens since the Ellis wormhole has a vanishing ADM mass.  
Abe found that microlensing light curves by the Ellis wormhole are dented near the peak of the light curves with a characteristic shape~\cite{Abe_2010}.

An extended source effect on light curves lensed by the Schwarzschild lens was investigated by Witt and Mao~\cite{Witt:1994}
and Nemiroff and Wickramasinghe~\cite{Nemiroff:1994uz} in 1994 and 
the first observed extended source effect was reported in 1997~\cite{Alcock:1997fi}.
Can we distinguish the light curve of an extended source microlensed by the Ellis wormhole from that by the Schwarzschild lens in their shapes? 
On this paper, we investigate an extended source effect on microlensing by the Ellis wormhole to answer this question.

This paper is organized as follows. 
In Sec.~II, we review the deflection angle of a light in the Ellis wormhole spacetime.
In Sec.~III, we investigate the total magnification of the images of the extended source lensed by the Ellis wormhole.
In Sec.~IV, we give a very short review for the Schwarzschild lens.
In Sec.~V, we consider microlensing of the extended source in the Ellis wormhole spacetime.
In Sec.~VI, we discuss and conclude our results.
In this paper we use the units in which the light speed and Newton's constant are unity.

\section{Deflection angle of a light in an Ellis wormhole spacetime}
In this section, we briefly review the deflection angle of a light~\cite{Chetouani_Clement_1984,Nandi_Zhang_Zakharov_2006,Muller:2008zza,Tsukamoto:2016qro,Tsukamoto:2016jzh,Jusufi:2017gyu} in an Ellis wormhole spacetime~\cite{Ellis_1973,Bronnikov_1973}.
A line element in the Ellis wormhole spacetime with a vanishing ADM mass is given by, 
in coordinates $-\infty < t < \infty$,   $-\infty < r < \infty$, $0 \leq  \theta \leq \pi$, and $0 \leq  \phi < 2\pi$, 
\begin{equation}
ds^2=-dt^2+dr^2+(r^2+a^2)(d\theta^2+\sin^2\theta d\phi^2),
\end{equation}
where $a$ is a positive constant. The throat of the wormhole is at $r=0$. 
We call $a$ the radius of the wormhole throat 
since a throat surface area is given by $4\pi a^2$.  
There exist a time translational Killing vector $t^\alpha \partial_\alpha =t^t$ 
and an axial Killing vector $\phi^\alpha \partial_\alpha =\phi^\phi$    
because of stationarity and axisymmery of the Ellis wormhole spacetime, respectively. 
We concentrate on a region $0 < r < \infty$ 
since we are interested in light rays which do not pass through the throat.

From an equation $k^\mu k_\mu=0$, where $k^\mu$ is the wave number of a photon, 
the trajectory equation of a light is given by 
\begin{equation}\label{eq:trajectory}
\frac{1}{(r^2+a^2)^2} \left( \frac{dr}{d\phi} \right)^2 = \frac{1}{b^2} -\frac{1}{r^2+a^2},
\end{equation}
where $b\equiv L/E$ is the impact parameter and $E\equiv -g_{\mu\nu}t^\mu k^\nu \geq 0$ and $L\equiv g_{\mu\nu}\phi^\mu k^\nu$ 
are the conserved energy and angular momentum of the photon, respectively.
Here we have assumed $\theta=\pi/2$ without loss of generality since the Ellis wormhole spacetime is a spherical spacetime.
A light ray is scattered when $\left| b \right| > a$ while it falls into the throat when $\left| b \right| \leq a$.
We concentrate on the scattered case with $\left| b \right| > a$. 
From Eq.~(\ref{eq:trajectory}), the absolute value of the deflection angle $\alpha$ of a light is obtained as
\begin{eqnarray}\label{deflection}
\left| \alpha \right|
&=& 2 \int^\infty_{r_0} \frac{\left| b \right|dr}{\sqrt{(r^2+a^2)(r^2+a^2-b^2)}}-\pi \nonumber\\
&=&2K(k)-\pi,
\end{eqnarray}
where $r_0\equiv \sqrt{b^2-a^2}$ is the closest distance of the light ray from the wormhole throat 
and $K(k)$ is the complete elliptic integral of the first kind defined as
\begin{equation}\label{first}
K(k)\equiv \int^1_0 \frac{dx}{\sqrt{(1-x^2)(1-k^2x^2)}},
\end{equation}
where $0<k\equiv a/\left| b \right|<1$ and we have used $x\equiv \left| b \right|/\sqrt{r^2+a^2}$.
Under a weak-field approximation $a \ll \left| b \right|$, we obtain the deflection angle as
\begin{equation}\label{eq:deflection_weak}
\alpha= \pm \frac{\pi}{4} \left( \frac{a}{b} \right)^2 +O\left(\left( \frac{a}{b} \right)^4\right).
\end{equation}
Here the upper (lower) sign is chosen if $b$ is positive (negative). 

The Ellis wormhole spacetime has a light sphere at $r=0$ which is coincidence to the throat. 
Notice that the deflection angle of the light $\alpha$ diverges in a strong deflection limit $\left| b \right|\rightarrow a$ or $k \rightarrow 1$ and 
that the light winds around the throat or the light sphere infinite times in the strong deflection limit.
See Ref.~\cite{Tsukamoto:2016qro} for the details of the deflection angle of the light in the strong deflection limit 
and under the weak-field approximation in the Ellis wormhole spacetime.

\section{Magnifications of the lensed images of an extended source}
\begin{figure}[htbp]
\begin{center}
\includegraphics[width=70mm]{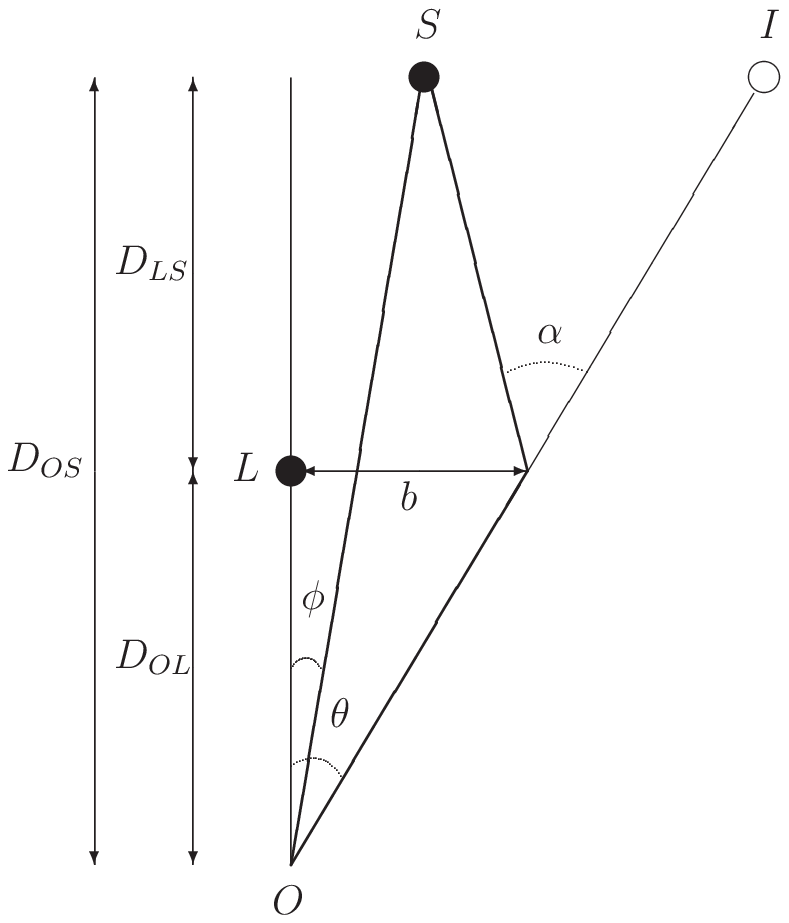}
\end{center}
\caption{Lens configuration of gravitational lensing. A light ray emitted by a source~$S$ with a source angle $\phi$ 
is deflected near a wormhole as a lens~$L$ with a deflection angle $\alpha$.
An observer $O$ sees an image~$I$ of the light ray with an image angle $\theta$.
$D_{OS}$, $D_{OL}$, and $D_{LS}$ are distances 
between the observer~$O$ and the source~$S$, between the observer~$O$ and the lens~$L$, and between the lens~$L$ and the source~$S$, respectively.
$b$ is the impact parameter of the light ray.
}
\label{Fig:lens_configuration}
\end{figure}

In this section, we investigate the magnifications of the images of an extended source lensed by an Ellis wormhole.
We consider that a light ray emitted by a source with a source angle $\phi$ is deflected by the Ellis wormhole as a lens with a deflection angle $\alpha$ 
and that an observer sees the image of the light ray with an image angle $\theta$.
We assume that all the angles are small, i.e., $\left| \alpha \right| \ll 1$, $\left| \theta \right| \ll 1$, and $\left| \phi \right| \ll 1$.
Under the assumptions, a lens equation is obtained as
\begin{equation}\label{eq:lens}
D_{LS}\alpha=D_{OS}(\theta-\phi),
\end{equation}
where $D_{LS}$ is the distance between the lens and the source and $D_{OS}$ is the distance between the observer and the source. 
Note that the distance $D_{OL}$ between the observer and the lens is given by $D_{OL}=D_{OS}-D_{LS}$ 
and that the impact parameter of the light ray $b$ is given by $b=D_{OL}\theta$ under the assumptions.
Figure~\ref{Fig:lens_configuration} shows the lens configuration of gravitational lensing.

From the deflection angle under the weak-field approximation~(\ref{eq:deflection_weak}), 
we can express the lens equation~(\ref{eq:lens}) as
\begin{equation}\label{eq:lens_2}
\pm \hat{\theta}^{-2}=\hat{\theta}-\hat{\phi},
\end{equation}
where the upper (lower) sign is chosen if $\hat{\theta}$ or $b$ is positive (negative), 
$\hat{\theta}\equiv \theta/\theta_\mathrm{E}$ and $\hat{\phi}\equiv \phi/\theta_\mathrm{E}$, 
and $\theta_\mathrm{E}$ is the image angle of an Einstein ring given by
\begin{equation}\label{eq:Einstein}
\theta_\mathrm{E} \equiv \left( \frac{\pi a^2D_{LS}}{4D_{OS}D_{OL}^2} \right)^\frac{1}{3}.
\end{equation}
The lens equation~(\ref{eq:lens_2}) has only a positive solution $\hat{\theta}=\hat{\theta}_+$ 
and a negative solution $\hat{\theta}=\hat{\theta}_-$ for any given $\hat{\phi}$. 
We notice that the lens equation~(\ref{eq:lens_2}) has point symmetry against a point $\hat{\theta}=\hat{\phi}=0$.
From point symmetry, we obtain
\begin{equation}
\hat{\theta}_-(\hat{\phi})=-\hat{\theta}_+(-\hat{\phi})
\end{equation}
and then 
\begin{equation}
\mu_{\mathrm{p}-}(\hat{\phi})=\mu_{\mathrm{p}+}(-\hat{\phi}), 
\end{equation}
where $\mu_{\mathrm{p} \pm}(\hat{\phi})$ is the magnifications of the images of a point source 
with the reduced image angles $\hat{\theta}_\pm(\hat{\phi})$ defined as
\begin{equation}\label{eq:magnification_point}
\mu_{\mathrm{p} \pm}(\hat{\phi})\equiv \frac{\hat{\theta}_\pm(\hat{\phi})}{\hat{\phi}} \frac{d\hat{\theta}_\pm}{d\hat{\phi}}(\hat{\phi}). 
\end{equation}
We can concentrate on the case for $\hat{\phi}\geq 0$ without loss of generality because of point symmetry.
From the lens equation~(\ref{eq:lens_2}), 
the magnifications of the images of the point source $\mu_{\mathrm{p} \pm}$ are rewritten as
\begin{equation}
\mu_{\mathrm{p}\pm}(\hat{\phi})=\frac{\hat{\theta}_{\pm}^6(\hat{\phi})}{\left( \hat{\theta}_{\pm}^3(\hat{\phi}) \mp 1 \right) \left( \hat{\theta}_{\pm}^3(\hat{\phi}) \pm 2 \right)}.
\end{equation}

The total magnification of the images of the point source $\mu_\mathrm{p}$ is defined as 
\begin{equation}\label{eq:total_magnification_point}
\mu_\mathrm{p}(\hat{\phi}) \equiv \left| \mu_{\mathrm{p}+}(\hat{\phi}) \right| +\left| \mu_{\mathrm{p}-}(\hat{\phi}) \right|. 
\end{equation}
In the region $1\ll \hat{\phi} \ll \theta_\mathrm{E}^{-1}$,
the total magnification of the point source $\mu_\mathrm{p}$ is given by~\cite{Tsukamoto:2016zdu}
\begin{equation}\label{eq:total_magnification_point_edge}
\mu_\mathrm{p}\sim 1-\frac{1}{2\hat{\phi}^3}<1.
\end{equation}

The total magnification of the images of an extended source $\mu_\mathrm{e}$ is obtained as~\cite{Schneider_Ehlers_Falco_1992}
\begin{equation}\label{eq:total_magnification0}
\mu_\mathrm{e}(\hat{\mbox{\boldmath $\phi$}})
=\frac{\iint d\hat{\phi}'^2 I(\hat{\mbox{\boldmath $\phi$}}')\mu_\mathrm{p}(\hat{\mbox{\boldmath $\phi$}}')}{\iint d\hat{\phi}'^2 I(\hat{\mbox{\boldmath $\phi$}}')}, 
\end{equation}
where $I(\hat{\mbox{\boldmath $\phi$}}')$ is a surface brightness profile of the extended source 
and $\hat{\mbox{\boldmath $\phi$}}'$ is a normalized position vector on a source plane
in units of $R_\mathrm{E}$.
Here $R_\mathrm{E}$ is the radius of the Einstein ring on the source plane defined as 
\begin{equation}\label{eq:R_E}
R_\mathrm{E} \equiv D_{OS}\theta_\mathrm{E}.
\end{equation}

We consider an extended source with a radius $R_S$ and with a uniform surface brightness.
Introducing the dimensionless radius of the extended source on the source plane $\hat{\phi}_S\equiv R_S/R_\mathrm{E}$,
the total magnification of the images of the extended source $\mu_\mathrm{e}$ is given by
\begin{equation}\label{eq:total_magnification1}
\mu_\mathrm{e}(\hat{\phi})
=\frac{2}{\pi \hat{\phi}_S^2} \int^{\hat{\phi}+\hat{\phi}_S}_{\hat{\phi}-\hat{\phi}_S} \mu_\mathrm{p}(\hat{\phi}')\hat{\phi}' \arccos \frac{\hat{\phi}'^2+\hat{\phi}^2-\hat{\phi}_S^2}{2\hat{\phi}\hat{\phi}'} d\hat{\phi}' 
\end{equation}
for $\hat{\phi}_S<\hat{\phi}$
and 
\begin{eqnarray}\label{eq:total_magnification2}
\mu_\mathrm{e}(\hat{\phi})
&=&\frac{2}{\pi \hat{\phi}_S^2} \int^{\hat{\phi}+\hat{\phi}_S}_{-\hat{\phi}+\hat{\phi}_S} \mu_\mathrm{p}(\hat{\phi}')\hat{\phi}' \arccos \frac{\hat{\phi}'^2+\hat{\phi}^2-\hat{\phi}_S^2}{2\hat{\phi}\hat{\phi}'} d\hat{\phi}' \nonumber\\
&&+\frac{2}{\hat{\phi}_S^2} \int^{-\hat{\phi}+\hat{\phi}_S}_{0}  \mu_\mathrm{p}(\hat{\phi}')\hat{\phi}' d\hat{\phi}'
\end{eqnarray}
for $\hat{\phi}<\hat{\phi}_S$.
In a perfect aligned case with $\hat{\phi}=0$, we obtain the total magnification as 
\begin{eqnarray}\label{eq:total_magnification3}
\mu_\mathrm{e}(0)
=\frac{2}{\hat{\phi}_S^2} \int^{\hat{\phi}_S}_{0}  \mu_\mathrm{p}(\hat{\phi}')\hat{\phi}' d\hat{\phi}'.
\end{eqnarray}

\section{Schwarzschild Lens}
In this section, we give a very short review for a Schwarzschild lens with a positive ADM mass $M$ under a weak-field approximation~\cite{Schneider_Ehlers_Falco_1992}.
The deflection angle of a light is given by 
\begin{equation}\label{eq:deflection_mass}
\alpha =\frac{4M}{b}.
\end{equation}
Using the deflection angle~(\ref{eq:deflection_mass}), 
the lens equation~(\ref{eq:lens}) is rewritten as 
\begin{equation}\label{eq:lens_mass}
\hat{\theta}^{-1}=\hat{\theta}-\hat{\phi}.
\end{equation}
The lens equation has only a positive and a negative solutions for any reduced source angle $\hat{\phi}$.
Note that the lens equation (\ref{eq:lens_mass}) has point symmetry against a point $\hat{\theta}=\hat{\phi}=0$.
We can concentrate ourselves on the case for $\hat{\phi} \geq 0$ without loss of generality because of point symmetry.

The image angle of an Einstein ring is obtained as 
\begin{equation}\label{eq:Einstein_mass}
\theta_\mathrm{E}=\sqrt{\frac{4MD_{LS}}{D_{OS}D_{OL}}}.
\end{equation}
From Eqs.~(\ref{eq:Einstein}) and (\ref{eq:Einstein_mass}), 
we notice that the angle of the Einstein ring $\theta_\mathrm{E}$ by the Schwarzschild lens 
is the same as the one by the Ellis wormhole with the radius of the throat $a$
when the ADM mass is
\begin{equation}\label{eq:fine-tuned mass}
M=\frac{1}{8}\left( \frac{\pi^2a^4D_{OS}}{2D_{LS}D_{OL}} \right)^\frac{1}{3}.
\end{equation}

The total magnification of the images of a point source is obtained as
\begin{equation}\label{eq:total_magnification_point_mass}
\mu_\mathrm{p}(\hat{\phi})= \frac{\hat{\phi}^2+2}{\hat{\phi}\sqrt{\hat{\phi}^2+4}}. 
\end{equation}
In the region $1\ll \hat{\phi} \ll \theta_\mathrm{E}^{-1}$,
the total magnification of the point source $\mu_\mathrm{p}$ is given by
\begin{equation}\label{eq:total_magnification_point_edge_mass}
\mu_\mathrm{p}\sim 1+\frac{2}{\hat{\phi}^4}>1.
\end{equation}
The total magnification of the images of an extended source $\mu_\mathrm{e}(\hat{\phi})$ with a uniform surface brightness 
is given by Eqs.~(\ref{eq:total_magnification1}) and (\ref{eq:total_magnification2}) with Eq.~(\ref{eq:total_magnification_point_mass}).

\section{Microlensing by an Ellis wormhole}
\begin{figure}[htbp]
\begin{center}
\includegraphics[width=70mm]{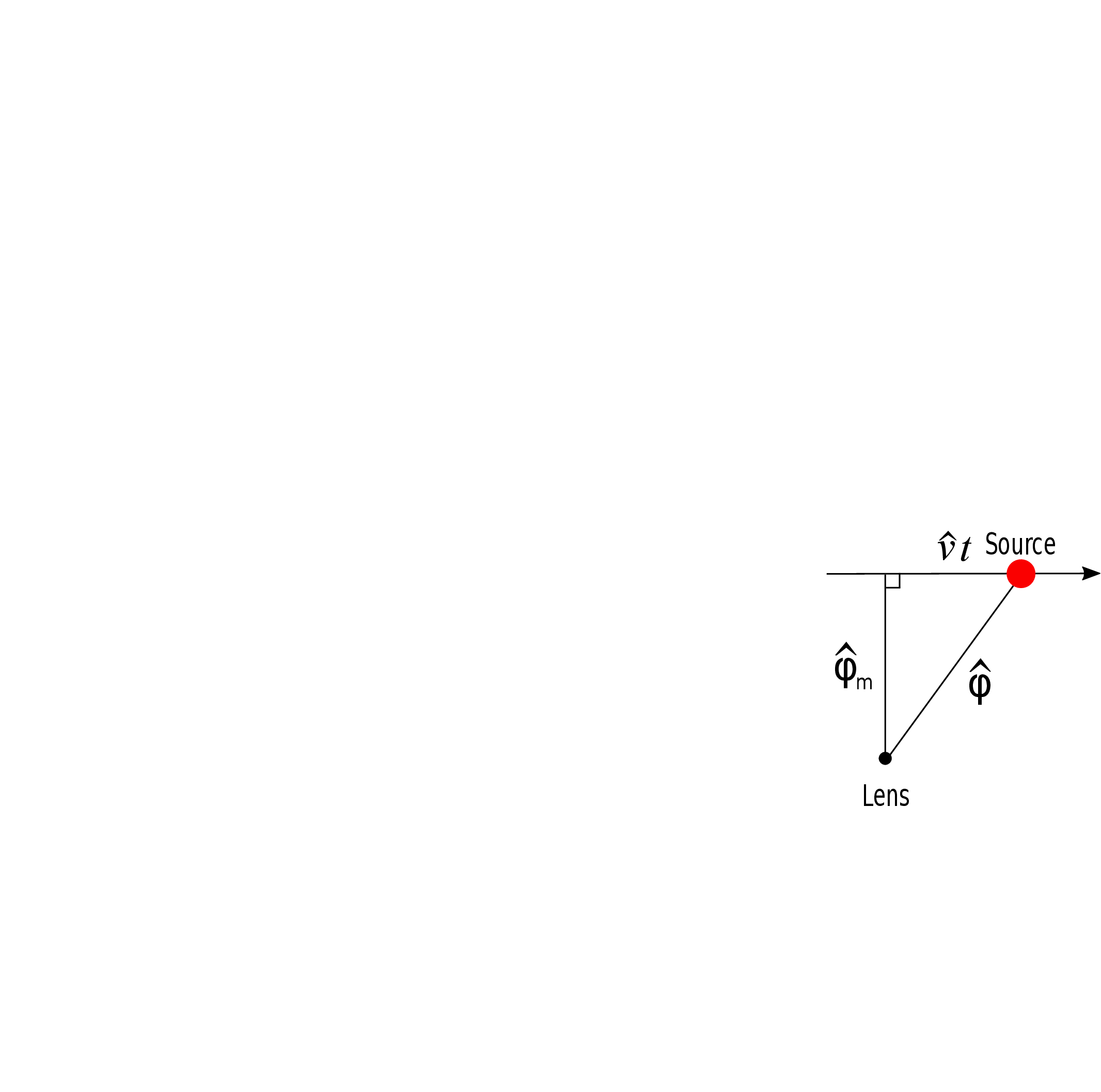}
\end{center}
\caption{The motion of an extended source on a source plane in units of $R_\mathrm{E}$. 
The reduced angle $\hat{\phi}$ defined at the location of an observer between the center of the extended source and an optical axis $\hat{\phi}=0$ 
is described by $\hat{\phi}=\sqrt{\hat{\phi}_\mathrm{m}^2+\hat{v}^2t^2}$, 
where $\hat{\phi}_\mathrm{m}$ is the closest reduced angle,
$\hat{v}\equiv v/R_\mathrm{E}$ is the relative velocity of the source against the optical axis on the source plane, 
and $t$ is a time.
We have set the time $t$ to be $t=0$ when $\hat{\phi}=\hat{\phi}_\mathrm{m}$.
}
\label{Fig:source_plane}
\end{figure}

We consider the microlensing~\cite{Paczynski_1986} of an extended source which moves with a relative velocity $v$ against an optical axis $\hat{\phi}=0$ 
and which is lensed by an Ellis wormhole. 
An Einstein-ring-radius-crossing time $t_\mathrm{E}$ defined as $t_\mathrm{E}\equiv R_\mathrm{E}/v$ gives the timescale of the microlensing.
See Fig.~\ref{Fig:source_plane} for the motion of the source.
We have set a time $t$ to be $t=0$ when the reduced angle of the center of the extended source $\hat{\phi}=\hat{\phi}_\mathrm{m}$, 
where $\hat{\phi}_\mathrm{m}$ is the closest reduced angle.
In the wide range of the radius of the throat $a$, 
the angle of the Einstein ring $\theta_\mathrm{E}$,
the radius of the Einstein ring on the source plane $R_\mathrm{E}$, 
and the Einstein-ring-radius-crossing time $t_\mathrm{E}$ are shown in Tables~\ref{table:I} and \ref{table:II}. 
\begin{table}[hbtp]
 \caption{Microlensing of a star at $D_{OS}=8$~kpc in the bulge of our galaxy lensed by an Ellis wormhole at $D_{OL}=4$~kpc. 
 We assume that the relative velocity $v$ of the star against an optical axis on a source plane is $v=200$~km/s.}
 \label{table:I}
\begin{center}
\begin{tabular}{c c c c} \hline
$a$(km) &$\theta_\mathrm{E}$(mas) &$R_\mathrm{E}$(km) &$t_\mathrm{E}$(day) \\ \hline
$1.0\times 10^{-2}      \quad   $ & $ 2.8\times 10^{-5}    \quad   $ & $ 3.4\times 10^{4}    \quad   $ & $ 2.0\times 10^{-3}$ \\ 
$1.0\times 10^{-1}      \quad   $ & $ 1.3\times 10^{-4}    \quad   $ & $ 1.6\times 10^{5}    \quad   $ & $ 9.1\times 10^{-3}$ \\ 
$1.0                    \quad   $ & $ 6.1\times 10^{-4}    \quad   $ & $ 7.3\times 10^{5}    \quad   $ & $ 4.2\times 10^{-2}$ \\ 
$1.0\times 10^{1}       \quad   $ & $ 2.8\times 10^{-3}    \quad   $ & $ 3.4\times 10^{6}    \quad   $ & $ 2.0\times 10^{-1}$ \\ 
$1.0\times 10^{2}       \quad   $ & $ 1.3\times 10^{-2}    \quad   $ & $ 1.6\times 10^{7}    \quad   $ & $ 9.1\times 10^{-1}$ \\ 
$1.0\times 10^{3}       \quad   $ & $ 6.1\times 10^{-2}    \quad   $ & $ 7.3\times 10^{7}    \quad   $ & $ 4.2$               \\ 
$1.0\times 10^{4}       \quad   $ & $ 2.8\times 10^{-1}    \quad   $ & $ 3.4\times 10^{8}    \quad   $ & $ 2.0\times 10^{1}$  \\ 
$1.0\times 10^{5}       \quad   $ & $ 1.3                  \quad   $ & $ 1.6\times 10^{9}    \quad   $ & $ 9.1\times 10^{1}$  \\ 
$1.0\times 10^{6}       \quad   $ & $ 6.1                  \quad   $ & $ 7.3\times 10^{9}    \quad   $ & $ 4.2\times 10^{2}$  \\ 
$1.0\times 10^{7}       \quad   $ & $ 2.8\times 10^{1}     \quad   $ & $ 3.4\times 10^{10}   \quad   $ & $ 2.0\times 10^{3}$  \\ 
$1.0\times 10^{8}       \quad   $ & $ 1.3\times 10^{2}     \quad   $ & $ 1.6\times 10^{11}   \quad   $ & $ 9.1\times 10^{3}$  \\ 
$1.0\times 10^{9}       \quad   $ & $ 6.1\times 10^{2}     \quad   $ & $ 7.3\times 10^{11}   \quad   $ & $ 4.2\times 10^{4}$  \\ 
$1.0\times 10^{10}      \quad   $ & $ 2.8\times 10^{3}     \quad   $ & $ 3.4\times 10^{12}   \quad   $ & $ 2.0\times 10^{5}$  \\ \hline
\end{tabular}
\end{center}
\end{table}
\begin{table}[hbtp]
 \caption{Microlensing of a star at $D_{OS}=50$~kpc in the Large Magellanic Cloud lensed by an Ellis wormhole at $D_{OL}=25$~kpc.
 We assume that the relative velocity $v$ of the star against an optical axis on a source plane is $v=200$~km/s.}
 \label{table:II}
\begin{center}
\begin{tabular}{c c c c} \hline
$a$(km) &$\theta_\mathrm{E}$(mas) &$R_\mathrm{E}$(km) &$t_\mathrm{E}$(day) \\ \hline
$1.0\times 10^{-2}       \quad   $ & $ 8.3\times 10^{-6}    \quad   $ & $ 6.2\times 10^{4}    \quad   $ & $ 3.6\times 10^{-3}$ \\ 
$1.0\times 10^{-1}       \quad   $ & $ 3.9\times 10^{-5}    \quad   $ & $ 2.9\times 10^{5}    \quad   $ & $ 1.7\times 10^{-2}$ \\ 
$1.0                     \quad   $ & $ 1.8\times 10^{-4}    \quad   $ & $ 1.3\times 10^{6}    \quad   $ & $ 7.8\times 10^{-2}$ \\ 
$1.0\times 10^{1}        \quad   $ & $ 8.3\times 10^{-4}    \quad   $ & $ 6.2\times 10^{6}    \quad   $ & $ 3.6\times 10^{-1}$ \\ 
$1.0\times 10^{2}        \quad   $ & $ 3.9\times 10^{-3}    \quad   $ & $ 2.9\times 10^{7}    \quad   $ & $ 1.7$               \\ 
$1.0\times 10^{3}        \quad   $ & $ 1.8\times 10^{-2}    \quad   $ & $ 1.3\times 10^{8}    \quad   $ & $ 7.8$               \\ 
$1.0\times 10^{4}        \quad   $ & $ 8.3\times 10^{-2}    \quad   $ & $ 6.2\times 10^{8}    \quad   $ & $ 3.6\times 10^{1}$  \\ 
$1.0\times 10^{5}        \quad   $ & $ 3.9\times 10^{-1}    \quad   $ & $ 2.9\times 10^{9}    \quad   $ & $ 1.7\times 10^{2}$  \\ 
$1.0\times 10^{6}        \quad   $ & $ 1.8                  \quad   $ & $ 1.3\times 10^{10}   \quad   $ & $ 7.8\times 10^{2}$  \\ 
$1.0\times 10^{7}        \quad   $ & $ 8.3                  \quad   $ & $ 6.2\times 10^{10}   \quad   $ & $ 3.6\times 10^{3}$  \\ 
$1.0\times 10^{8}        \quad   $ & $ 3.9\times 10^{1}     \quad   $ & $ 2.9\times 10^{11}   \quad   $ & $ 1.7\times 10^{4}$  \\ 
$1.0\times 10^{9}        \quad   $ & $ 1.8\times 10^{2}     \quad   $ & $ 1.3\times 10^{12}   \quad   $ & $ 7.8\times 10^{4}$  \\ 
$1.0\times 10^{10}       \quad   $ & $ 8.3\times 10^{2}     \quad   $ & $ 6.2\times 10^{12}   \quad   $ & $ 3.6\times 10^{5}$  \\ \hline 
\end{tabular}
\end{center}
\end{table}

As Abe found in Ref.~\cite{Abe_2010}, the total magnification of a point source lensed by the Ellis wormhole can be smaller than unity 
when the point source is outside of the Einstein ring on the source plane.
The gutters of light curves of the point source microlensed by the Ellis wormhole appear near the peak of the light curves.
We can look for the Ellis wormholes by microlensing light curves with the gutters.

\begin{figure*}[htbp]
\begin{center}
\includegraphics[width=80mm]{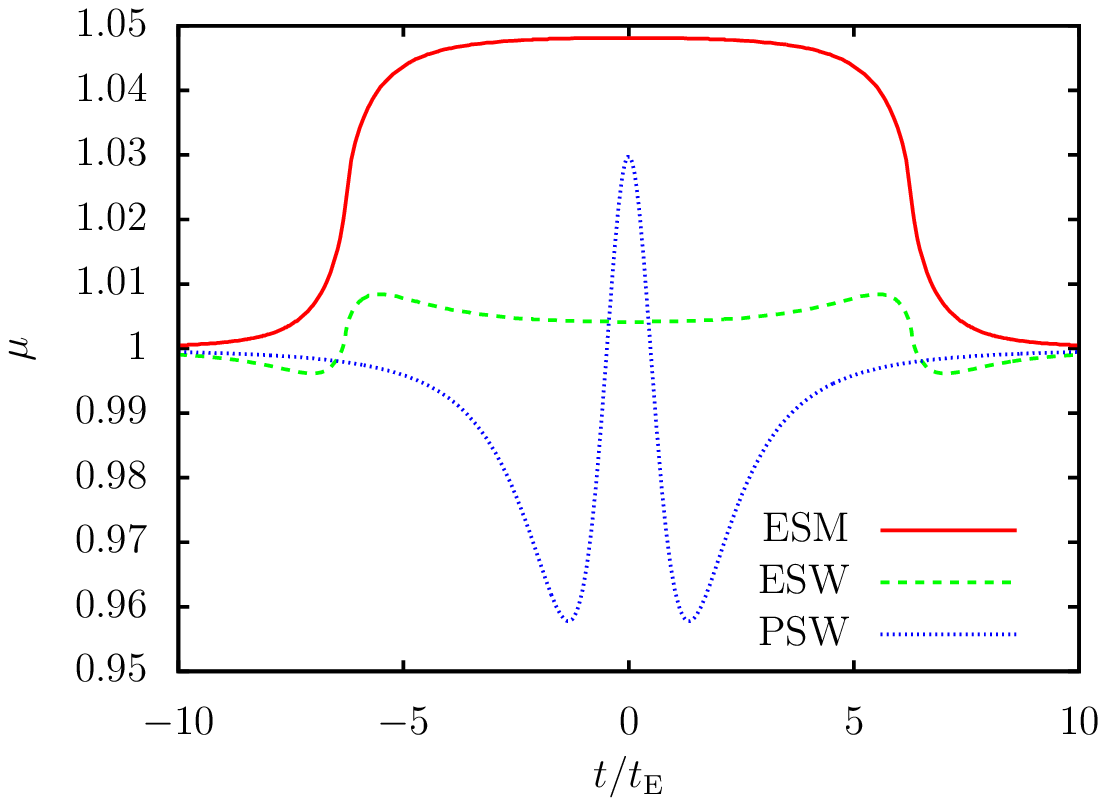}
\includegraphics[width=80mm]{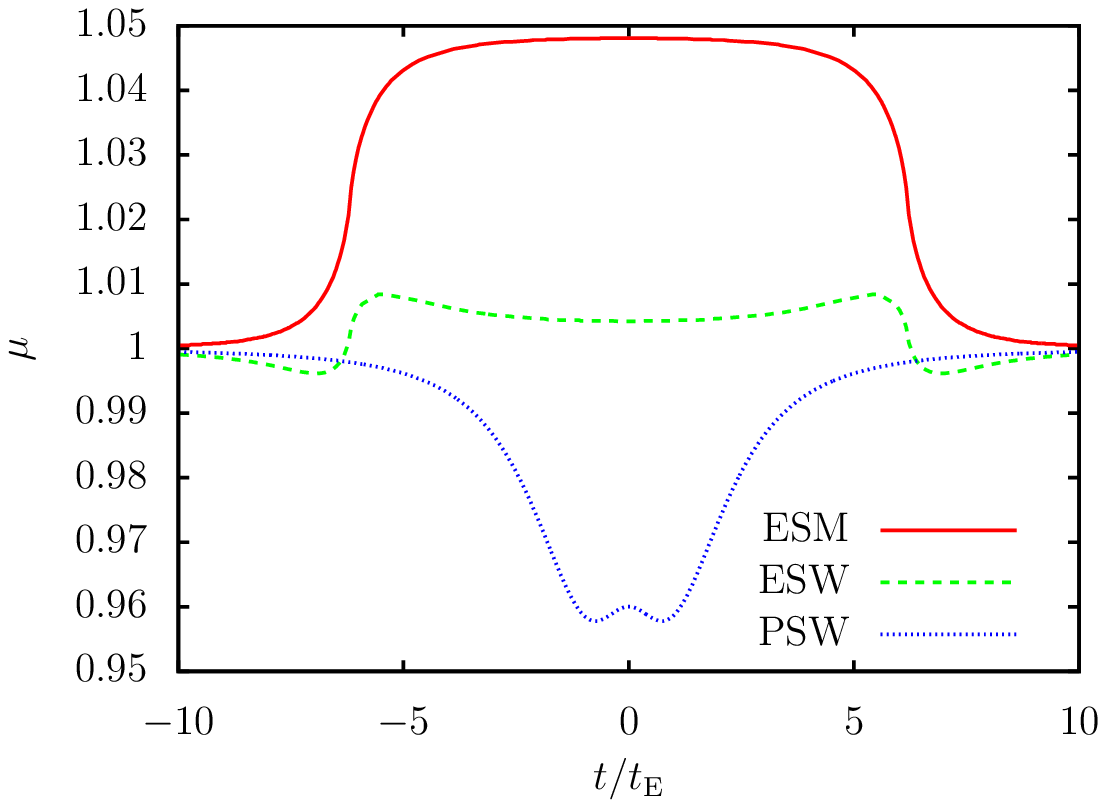}
\includegraphics[width=80mm]{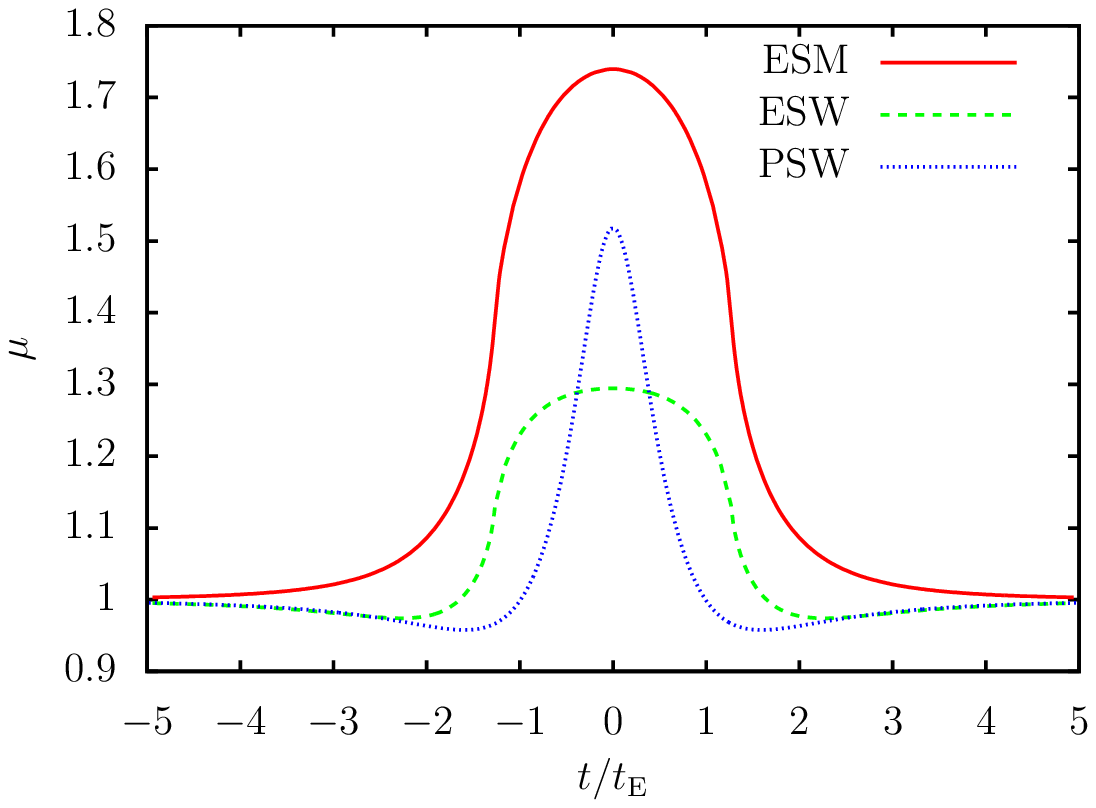}
\includegraphics[width=80mm]{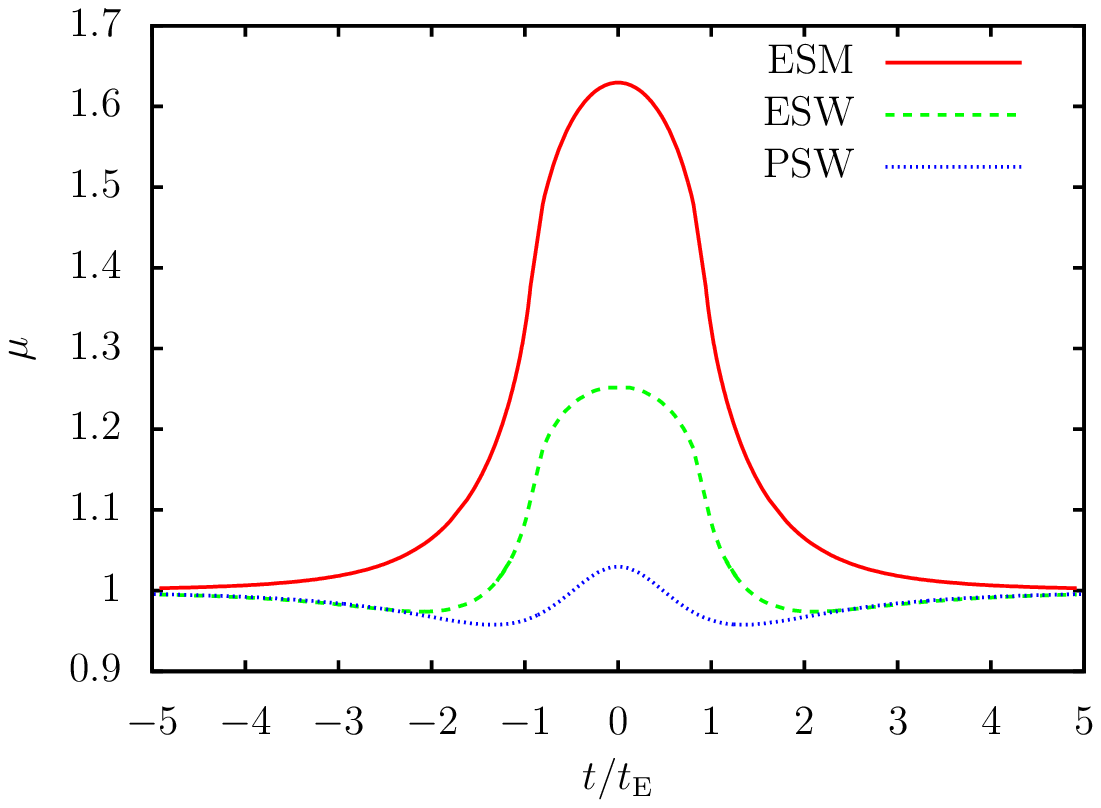}
\end{center}
\caption{Microlensing light curves with $D_{OL}=D_{LS}=4$~kpc and with $v=200$~km/s.
Solid-red, broken-green, and dotted-blue curves denote 
the light curves of an Extended Source with a radius $R_S=10^6$~km lensed by an Mass (ESM),
an Extended Source with $R_S=10^6$~km lensed by a massless Wormhole (ESW),
and a Point Source lensed by the massless Wormhole (PSW), respectively.
The \textit{left top} panel shows the light curves with $a=0.1$~km and the closest separation $\hat{\phi}_\mathrm{m}=1$, 
the \textit{right top} panel shows the light curves with $a=0.1$~km and $\hat{\phi}_\mathrm{m}=1.5$, 
the \textit{left bottom} panel shows the light curves with $a=1$~km and $\hat{\phi}_\mathrm{m}=0.5$, 
and the \textit{right bottom} panel shows the light curves with $a=1$~km and $\hat{\phi}_\mathrm{m}=1$.
The vertical axis $\mu$ denotes the total magnification 
and it is given by $\mu=\mu_\mathrm{e}$ in the ESM and ESW cases while $\mu=\mu_\mathrm{p}$ in the PSW case.
By using Eq.~(\ref{eq:fine-tuned mass}), we have tuned the ADM mass of the mass lens 
to have the same size of the Einstein ring $\theta_\mathrm{E}$ as the size of the Einstein ring made by the wormhole.
}
\label{Fig:light_curves}
\end{figure*}

If the radius of a source $R_S$ is larger than the radius of the Einstein ring $R_\mathrm{E}$ on the source plane, 
an extended source effect on microlensing light curves cannot be neglected. 
From Eqs.~(\ref{eq:Einstein}) and (\ref{eq:R_E}), we obtain the radius of the wormhole throat $a$ satisfying $R_\mathrm{E}=R_S$ as 
\begin{equation}\label{eq:critical_a}
a=\sqrt{\frac{4R_S^3D_{OL}^2}{\pi D_{OS}^2D_{LS}}}.
\end{equation}
Figure~\ref{Fig:light_curves} shows the examples of the light curves of a point source and an extended source with a radius $R_S=10^6$~km 
at $D_{OS}=$ 8~kpc microlensed by an Ellis wormhole at $D_{OL}=4$~kpc. 
In this case, the extended source effect cannot be neglected 
when the Ellis wormhole has the radius of the throat $a \lesssim 1.6$~km obtained from Eq.~(\ref{eq:critical_a}).
Thus, we consider light curves microlensed by Ellis wormholes with the radii of the throat $a=1$~km and $a=0.1$~km in Fig~\ref{Fig:light_curves}.  
As a reference, the light curves of the extended source lensed by a Schwarzschild lens at $D_{OL}=4$~kpc is also plotted there. 
By using Eq.~(\ref{eq:fine-tuned mass}), we have tuned the ADM mass of the Schwarzschild lens 
to have the same size of the Einstein ring $\theta_\mathrm{E}$ as the size of the Einstein ring made by the Ellis wormhole.

\section{Discussion and Conclusion}
We discuss the extended source effect on the light curves microlensed by the Ellis wormhole shown in Fig.~\ref{Fig:light_curves}.
For discussing it, we should keep in mind that the point source lensed by the Ellis wormhole is demagnified (magnified)
if the point source is in the outer side excluding the vicinity of (inside and in the vicinity of) the Einstein ring on the source plane. 
We notice that the light curves have a symmetry against $t=0$.
The shape of the light curves of the point source lensed by the Ellis wormhole and the Schwarzschild lens 
only depends on the the closest separation $\hat{\phi}_\mathrm{m}$ 
while the one of the extended source not only depends on $\hat{\phi}_\mathrm{m}$ 
but also the dimensionless radius of the extended source $\hat{\phi}_S$.

The timescale of microlensing is determined by the larger one between the crossing time of the Einstein ring radius $t_\mathrm{E}$ 
and the crossing time of the extended source $R_{S}/v$. 
If the timescale of microlensing is less than a day,
it is difficult to detect microlensing with current microlensing observations at the interval of a few hours.
We can detect microlensing of a star at $D_{OS}=8$~kpc in the bulge of our galaxy 
by the Ellis wormhole with the radius of the throat $a=1$~km at $D_{OL}=4$~kpc 
if the relative velocity $v$ of the star against an observer-lens system is smaller than $10$~km/s.

As shown on the \textit{left top} (\textit{left bottom}) panel in Fig.~\ref{Fig:light_curves}, 
the light curves of the extended source microlensed by the Ellis wormhole with the throat radius $a=0.1$~km ($a=1$~km) 
and with the closest separation $\hat{\phi}_\mathrm{m}=1$ ($\hat{\phi}_\mathrm{m}=0.5$) flatten
while the light curves of the point source microlensed by the Ellis wormhole have a peak and gutters.
The flattening of the peak and gutters of the light curves is caused by the cancellation between the demagnified part and the magnified part of the extended source.
Thus, the gutters of the light curves of the extended source lensed by the Ellis wormhole cannot be deeper 
than the gutters by approximately $4\%$ demagnification which is obtained in the point source case.

As shown on \textit{right top} panel,
the point source with $\hat{\phi}_\mathrm{m}=1.5$ does not across the Einstein ring on the source plane
and the light curve of the point source lensed by the Ellis wormhole with $a=0.1$~km 
is always demagnified. 
On the other hand, the light curve of the extended source lensed by the Ellis wormhole with $\hat{\phi}_\mathrm{m}=1.5$ 
is magnified for $\left| t \right| \lesssim R_{S}/v$   
since a part of the extended source is located at the magnified region on the source plane. 

As shown on the \textit{right bottom} panel,
the peak of the light curve of the extended source lensed by the Ellis wormhole with $a=1$~km and with $\hat{\phi}_\mathrm{m}=1.0$ 
is higher than the one of the point source
since a part of the extended source can be extended at the magnified region on the source plane. 

The extended source effect on the total magnification is tiny for the region $1\ll \hat{\phi} \ll \theta_\mathrm{E}^{-1}$.
Thus, the behavior of the edge of the light curves will be described well by, 
from Eqs.~(\ref{eq:total_magnification_point_edge}) and (\ref{eq:total_magnification_point_edge_mass}),  
\begin{equation}
\mu_\mathrm{e}\sim \mu_\mathrm{p}\sim 1-\frac{1}{2\hat{\phi}^3}<1
\end{equation}
and
\begin{equation}
\mu_\mathrm{e}\sim \mu_\mathrm{p}\sim 1+\frac{2}{\hat{\phi}^4}>1
\end{equation}
for the Ellis wormhole and the Schwarzschild lens, respectively. 
This implies that the light curves lensed by the Ellis wormhole do not coincide with the ones lensed by the Schwarzschild lens in general
even if the values of $\hat{\phi}_\mathrm{m}$ and $\hat{\phi}_S$ are changed.

Figure~\ref{Fig:light_curves} shows that
the total magnification of the images of the extended source 
lensed by the Schwarzschild lens monotonically increase as the time $t$ increases for a period $t\leq 0$ 
while the ones by the Ellis wormhole decrease and then increase or decrease, increase and then decrease for the period $t\leq 0$.

From the above, we conclude that we can distinguish between the light curves of the extended source lensed by the Ellis wormhole 
and the ones lensed by the Schwarzschild lens in their shapes
even if the size of the extended source is a few times larger than the size of the Einstein ring on the source plane. 

In the end, we comment on the separation of a couple of lensed images.
The size of the image separation is the same as the size of the Einstein ring approximately.
Table~I (II) shows that a couple of images lensed by an Ellis wormhole $4$ ($25$) kpc away from an observer with a throat which is larger than $a=10^4$ ($10^5$) km 
could be separated into two images by using Very Long Baseline Interferometer (VLBI) 
such as VLBI Exploration of Radio Astrometry~\cite{Kobayashi_2003} with the resolution of sub-milliarcsecond.

\section*{Acknowledgements}
The authors would like to thank Rajibul Shaikh for his useful and kind comments.
This work was supported in part by the National Natural Science Foundation of China under Grant No. 11475065 and
the Major Program of the National Natural Science Foundation of China under Grant No. 11690021.


\begin{thebibliography}{99}



\bibitem{Visser_1995} 
M. Visser, 
\textit{Lorentzian Wormholes: From Einstein to Hawking} (American Institute of Physics, Woodbury, NY, 1995).

\bibitem{Schneider_Ehlers_Falco_1992} 
P. Schneider, J. Ehlers, and E. E. Falco, 
\textit{Gravitational Lenses} (Springer-Verlag, Berlin, 1992).

\bibitem{Petters_Levine_Wambsganss_2001}
A. O. Petters, H. Levine, and J. Wambsganss,
\textit{Singularity Theory and Gravitational Lensing} (Birkhauser, Boston, 2001).

\bibitem{Schneider_Kochanek_Wambsganss_2006} 
P. Schneider, C. S. Kochanek, and J. Wambsganss,
\textit{Gravitational Lensing: Strong, Weak and Micro,
Lecture Notes of the 33rd Saas-Fee Advanced Course},
edited by G. Meylan, P. Jetzer, and P. North (Springer-Verlag, Berlin, 2006).

\bibitem{Perlick_2004_Living_Rev} 
V. Perlick, 
Living Rev. Relativity {\bf7}, 9 (2004).

\bibitem{Kim_Cho_1994} 
S. W. Kim and Y. M. Cho, 
in \textit{Evolution of the Universe and its Observational Quest} (Universal Academy Press, Tokyo, 1994), p. 353.

\bibitem{Cramer:1994qj} 
  J.~G.~Cramer, R.~L.~Forward, M.~S.~Morris, M.~Visser, G.~Benford, and G.~A.~Landis,
  Phys.\ Rev.\ D {\bf 51}, 3117 (1995).

\bibitem{Nandi_Zhang_Zakharov_2006} 
K. K. Nandi, Y. Z. Zhang, and A. V. Zakharov, 
Phys. Rev. D {\bf 74}, 024020 (2006). 
  
\bibitem{Rahaman:2007am} 
  F.~Rahaman, M.~Kalam and S.~Chakraborty,
  Chin.\ J.\ Phys.\  {\bf 45}, 518 (2007).

\bibitem{Tejeiro_Larranaga_2012} 
J. M. Tejeiro and E. A. Larranaga, 
Rom. J. Phys. {\bf 57}, 736 (2012).

\bibitem{Kuhfittig:2015sta} 
P.~K.~F.~Kuhfittig,
Scientific Voyage {\bf 2}, 1 (2016).

\bibitem{Nandi:2016ccg} 
  K.~K.~Nandi, A.~A.~Potapov, R.~N.~Izmailov, A.~Tamang, and J.~C.~Evans,
  Phys.\ Rev.\ D {\bf 93}, 104044 (2016).

\bibitem{Tsukamoto:2016zdu} 
  N.~Tsukamoto and T.~Harada,
  Phys.\ Rev.\ D {\bf 95}, 024030 (2017).

\bibitem{Nandi:2016uzg} 
  K.~K.~Nandi, R.~N.~Izmailov, A.~A.~Yanbekov, and A.~A.~Shayakhmetov,
  Phys.\ Rev.\ D {\bf 95}, 104011 (2017).

\bibitem{Sajadi:2016hko} 
  S.~N.~Sajadi and N.~Riazi,
  arXiv:1611.04343 [gr-qc].

\bibitem{Jusufi:2017mav} 
  K.~Jusufi and A.~\"{O}vg\"{u}n,
  Phys.\ Rev.\ D {\bf 97}, 024042 (2018).

\bibitem{Shaikh:2017zfl} 
  R.~Shaikh and S.~Kar,
  Phys.\ Rev.\ D {\bf 96}, 044037 (2017).

\bibitem{Goulart:2017iko} 
  P.~Goulart,
  Class.\ Quant.\ Grav.\  {\bf 35}, 025012 (2018).

\bibitem{Chetouani_Clement_1984} 
L. Chetouani and G. Cl\'{e}ment, 
Gen. Relativ. Gravit. {\bf 16}, 111 (1984).

\bibitem{Perlick_2004_Phys_Rev_D} 
V. Perlick, 
Phys. Rev. D {\bf 69}, 064017 (2004). 

\bibitem{Muller:2008zza} 
T.~Muller,
Phys.\ Rev.\ D {\bf 77}, 044043 (2008);
T. K. Dey and S. Sen, 
Mod. Phys. Lett. A {\bf 23}, 953 (2008);
  A.~Bhattacharya and A.~A.~Potapov,
  Mod.\ Phys.\ Lett.\ A {\bf 25}, 2399 (2010);
G.~W.~Gibbons and M.~Vyska,
Class.\ Quant.\ Grav.\  {\bf 29}, 065016 (2012);
K. Nakajima and H. Asada, 
Phys. Rev. D {\bf 85}, 107501 (2012).

\bibitem{Abe_2010} 
F. Abe, 
Astrophys. J. {\bf 725}, 787 (2010). 

\bibitem{Toki_Kitamura_Asada_Abe_2011} 
Y. Toki, T. Kitamura, H. Asada, and F. Abe, 
Astrophys. J. {\bf 740}, 121 (2011). 

\bibitem{Tsukamoto_Harada_Yajima_2012} 
N. Tsukamoto, T. Harada, and K. Yajima,
Phys. Rev. D {\bf 86}, 104062 (2012).

\bibitem{Tsukamoto_Harada_2013} 
N. Tsukamoto and T. Harada,
Phys. Rev. D {\bf 87}, 024024 (2013). 

\bibitem{Yoo_Harada_Tsukamoto_2013}
C. M. Yoo, T. Harada, and N. Tsukamoto,
Phys. Rev. D {\bf 87}, 084045 (2013).

\bibitem{Takahashi_Asada_2013} 
R. Takahashi and H. Asada,
Astrophys. J. {\bf 768}, L16 (2013).

\bibitem{Izumi_2013} 
K.~Izumi, C.~Hagiwara, K.~Nakajima, T.~Kitamura, and H.~Asada,
Phys. Rev. D {\bf 88}, 024049 (2013).

\bibitem{Nakajima:2014nba} 
K.~Nakajima, K.~Izumi, and H.~Asada,
Phys.\ Rev.\ D {\bf 90}, 084026 (2014).

\bibitem{Bozza:2015haa} 
V.~Bozza and A.~Postiglione,
JCAP {\bf 1506}, 036 (2015).

\bibitem{Bozza:2015wbw} 
V.~Bozza and C.~Melchiorre,
JCAP {\bf 1603}, 040 (2016).
  
\bibitem{Lukmanova_2016}
R.~Lukmanova, A.~Kulbakova, R.~Izmailov, and A.~A.~Potapov, 
Int. J. Theor. Phys. {\bf 55}, 4723 (2016). 

\bibitem{Tsukamoto:2016jzh} 
  N.~Tsukamoto,
  Phys.\ Rev.\ D {\bf 95}, 064035 (2017).

\bibitem{Tsukamoto:2016qro} 
  N.~Tsukamoto,
  Phys.\ Rev.\ D {\bf 94}, 124001 (2016).

\bibitem{Jusufi:2017gyu} 
  K.~Jusufi,
  Int.\ J.\ Geom.\ Meth.\ Mod.\ Phys.\  {\bf 14}, 1750179 (2017).

\bibitem{Tsukamoto:2017edq} 
  N.~Tsukamoto,
  Phys.\ Rev.\ D {\bf 95}, 084021 (2017).

\bibitem{Jusufi:2017vta} 
  K.~Jusufi, A.~\"{O}vg\"{u}n, and A.~Banerjee,
  Phys.\ Rev.\ D {\bf 96}, 084036 (2017).

\bibitem{Bozza:2017dkv} 
  V.~Bozza,
  Int.\ J.\ Mod.\ Phys.\ D {\bf 26}, 1741013 (2017).

\bibitem{Asada:2017vxl} 
  H.~Asada,
  Mod.\ Phys.\ Lett.\ A {\bf 32}, 1730031 (2017).

\bibitem{Torres:1998xd} 
  D.~F.~Torres, G.~E.~Romero, and L.~A.~Anchordoqui,
  Phys.\ Rev.\ D {\bf 58}, 123001 (1998);
  L.~A.~Anchordoqui, G.~E.~Romero, D.~F.~Torres, and I.~Andruchow,
¡¦¡¦Mod.\ Phys.\ Lett.\ A {\bf 14}, 791 (1999);
  M.~Safonova, D.~F.~Torres, and G.~E.~Romero,
  Mod.\ Phys.\ Lett.\ A {\bf 16}, 153 (2001);
  E.~Eiroa, G.~E.~Romero, and D.~F.~Torres,
  Mod.\ Phys.\ Lett.\ A {\bf 16}, 973 (2001);
  M.~Safonova, D.~F.~Torres, and G.~E.~Romero,
  Phys.\ Rev.\ D {\bf 65}, 023001 (2001);
  M.~Safonova and D.~F.~Torres,
  Mod.\ Phys.\ Lett.\ A {\bf 17}, 1685 (2002).

\bibitem{Ellis_1973} 
H. G. Ellis, 
J. Math. Phys. {\bf 14}, 104 (1973). 

\bibitem{Bronnikov_1973}
K. A. Bronnikov, 
Acta Phys. Pol. B {\bf 4}, 251 (1973).

\bibitem{Morris_Thorne_1988} 
M. S. Morris and K. S. Thorne, 
Am. J. Phys. {\bf 56}, 395 (1988). 

\bibitem{Shinkai_Hayward_2002}
H. Shinkai and S. A. Hayward,
Phys. Rev. D {\bf 66}, 044005 (2002);
J. A. Gonz\'{a}lez, F. S. Guzm\'{a}n, and O. Sarbach
Class. Quant. Grav. {\bf 26}, 015010 (2009);
J. A. Gonz\'{a}lez, F. S. Guzm\'{a}n, and O. Sarbach
Class. Quant. Grav. {\bf 26}, 015011 (2009);
A. Doroshkevich, J. Hansen, I. Novikov, and A. Shatskiy,
Int. J. Mod. Phys. D {\bf 18}, 1665 (2009);
K.~A.~Bronnikov, J.~C.~Fabris, and A.~Zhidenko,
Eur.\ Phys.\ J.\ C {\bf 71}, 1791 (2011);
K.~A.~Bronnikov, R.~A.~Konoplya, and A.~Zhidenko,
Phys.\ Rev.\ D {\bf 86}, 024028 (2012).

\bibitem{Armendariz-Picon_2002}
C. Armend\'{a}riz-Pic\'{o}n,
Phys. Rev. D {\bf 65}, 104010 (2002).

\bibitem{Yazadjiev:2017twg} 
  S.~Yazadjiev,
  Phys.\ Rev.\ D {\bf 96}, 044045 (2017).

\bibitem{Kar:2002xa} 
S.~Kar, S.~SenGupta, and S.~Sur,
Phys.\ Rev.\ D {\bf 67}, 044005 (2003).

\bibitem{Das:2005un} 
A.~Das and S.~Kar,
Class.\ Quant.\ Grav.\  {\bf 22}, 3045 (2005).

\bibitem{Shatskiy:2008us} 
A.~Shatskiy, I.~D.~Novikov, and N.~S.~Kardashev,
Phys.\ Usp.\  {\bf 51}, 457 (2008).

\bibitem{Novikov:2012uj} 
I.~Novikov and A.~Shatskiy,
JETP {\bf 114}, 801 (2012).

\bibitem{Myrzakulov:2015kda} 
  R.~Myrzakulov, L.~Sebastiani, S.~Vagnozzi, and S.~Zerbini,
  Class.\ Quant.\ Grav.\  {\bf 33}, 125005 (2016).

\bibitem{Bronnikov:2013coa} 
K.~A.~Bronnikov, L.~N.~Lipatova, I.~D.~Novikov, and A.~A.~Shatskiy,
Grav.\ Cosmol.\  {\bf 19}, 269 (2013).  

\bibitem{Konoplya:2016hmd} 
  R.~A.~Konoplya and A.~Zhidenko,
  JCAP {\bf 1612}, 043 (2016).

\bibitem{Inada_Oguri_Shin_et_al_2012} 
N. Inada, M. Oguri, M. S. Shin \textit{et al.},
Astron. J. {\bf 143}, 119 (2012);
M. Oguri, N. Inada, B. Pindor \textit{et al.}, 
Astron. J. {\bf 132}, 999 (2006);
M. Oguri, N. Inada, M. A. Strauss \textit{et al.},
Astron. J. {\bf 135}, 512 (2008);
M. Oguri, N. Inada, M. A. Strauss \textit{et al.},
Astron. J. {\bf 143}, 120 (2012).

\bibitem{Barnacka_Glicenstein_Moderski_2012}
A. Barnacka, J.-F. Glicenstein, and M. Moderski,
Phys. Rev. D {\bf 86} 043001 (2012).

\bibitem{Meegan_Lichti_Bhat_et_al_2009} 
C. Meegan, G. Lichti, P. N. Bhat \textit{et al.},
Astrophys. J. {\bf 702}, 791 (2009). 

\bibitem{Muller_2004}
T.~Muller, 
Am. J. Phys. {\bf 72}, 1045,(2004).

\bibitem{Tsukamoto:2014swa} 
N.~Tsukamoto and C.~Bambi,
Phys.\ Rev.\ D {\bf 91}, 084013 (2015).

\bibitem{Ohgami:2015nra} 
T.~Ohgami and N.~Sakai,
Phys.\ Rev.\ D {\bf 91}, 124020 (2015). 

\bibitem{Ohgami:2016iqm} 
T.~Ohgami and N.~Sakai,
Phys.\ Rev.\ D {\bf 94}, 064071 (2016).

\bibitem{Perlick:2015vta} 
V.~Perlick, O.~Y.~Tsupko, and G.~S.~Bisnovatyi-Kogan,
Phys.\ Rev.\ D {\bf 92}, 104031 (2015).

\bibitem{Kitamura_Nakajima_Asada_2013} 
T. Kitamura, K. Nakajima, and H. Asada,
Phys. Rev. D {\bf 87}, 027501 (2013);
T.~Kitamura, K.~Izumi, K.~Nakajima, C.~Hagiwara, and H.~Asada,
Phys. Rev. D {\bf 89}, 084020 (2014);
  N.~Tsukamoto, T.~Kitamura, K.~Nakajima, and H.~Asada,
  Phys.\ Rev.\ D {\bf 90}, 064043 (2014).

\bibitem{Witt:1994}
H.~J.~Witt and S.~Mao, ApJ, {\bf 430}, 505 (1994).

\bibitem{Nemiroff:1994uz} 
  R.~J.~Nemiroff and W.~A.~D.~T.~Wickramasinghe,
  Astrophys.\ J.\  {\bf 424}, L21 (1994).
  
\bibitem{Alcock:1997fi} 
  C.~Alcock {\it et al.} [MACHO and GMAN Collaborations],
  Astrophys.\ J.\  {\bf 491}, 436 (1997).

\bibitem{Paczynski_1986} 
B.~Paczynski,
Astrophys. J. {\bf 304}, 1 (1986). 

\bibitem{Kobayashi_2003} 
H.~Kobayashi, {\it et al.}  in ASP Conf. Ser. 306, New
technologies in VLBI, ed. Y.C. Minh, (San Francisco:ASP), 367 (2003).



\end{thebibliography}
\end{document}